\documentclass[conference]{IEEEtran}
\IEEEoverridecommandlockouts

\usepackage{cite}
\usepackage{amsmath,amssymb,amsfonts}
\usepackage{algorithmic}
\usepackage{graphicx}
\usepackage{booktabs}

\usepackage{subcaption} 
\usepackage{algorithm}
\usepackage{algorithmic}

\usepackage{hyperref}
\usepackage{balance}

\usepackage{textcomp}
\usepackage{xcolor}
\usepackage{xspace}
\def\BibTeX{{\rm B\kern-.05em{\sc i\kern-.025em b}\kern-.08em
    T\kern-.1667em\lower.7ex\hbox{E}\kern-.125emX}}
\begin{document}

\newcommand{\ourname}{DLCodeGen\xspace}
\newcommand{\dlcodedataset}{DLCodeEval}
\newcommand{\solutiondataset}{DLPlanData}
\newcommand{\codebase}{DLSamplePool}

\definecolor{darkred}{rgb}{0.8, 0.1, 0.1}
\definecolor{darkgreen}{rgb}{0.1, 0.6, 0.1}
\definecolor{lightgray}{gray}{0.9}

\newcommand{\todoc}[2]{{\textcolor{#1}{\textbf{[#2]}}}}
\newcommand{\todoblue}[1]{\todoc{blue}{\textbf{#1}}}
\newcommand{\todored}[1]{\todoc{red}{\textbf{#1}}}
\newcommand{\tododarkred}[1]{\todoc{darkred}{\textbf{#1}}}

\newcommand{\xie}[1]{\todoblue{xie: #1}}
\newcommand{\gu}[1]{\todored{gu: #1}}
\newcommand{\shen}[1]{\tododarkred{shen: #1}}

\title{\fontsize{23.9pt}{30pt}\selectfont 
Empowering AI to Generate Better AI Code: Guided Generation of Deep Learning Projects with LLMs}

\author{\IEEEauthorblockN{Chen Xie, Mingsheng Jiao, Xiaodong Gu, Beijun~Shen$^{\dagger}$}\\
\IEEEauthorblockA{\textit{School of Computer Science, Shanghai Jiao Tong University, Shanghai, China}\\
\{dindin\_xc, jiaomingsheng, xiaodong.gu, bjshen\}@sjtu.edu.cn}}

\maketitle

\begin{abstract}
 While large language models (LLMs) have been widely applied to code generation, they struggle with generating entire deep learning projects, which are characterized by complex structures, longer functions, and stronger reliance on domain knowledge than general-purpose code. An open-domain LLM often lacks coherent contextual guidance and domain expertise for specific projects, making it challenging to produce complete code that fully meets user requirements. 
 In this paper, we propose a novel planning-guided code generation method, \ourname, tailored for generating deep learning projects. \ourname predicts a structured solution plan, offering global guidance for LLMs to generate the project. The generated plan is then leveraged to retrieve semantically analogous code samples and subsequently abstract a code template. To effectively integrate these multiple retrieval-augmented techniques, a comparative learning mechanism is designed to generate the final code. We validate the effectiveness of our approach on a dataset we build for deep learning code generation. Experimental results demonstrate that \ourname outperforms other baselines, achieving improvements of 9.7\% in CodeBLEU and 3.6\% in human evaluation metrics.
\end{abstract}


\section{Introduction}
$\let\thefootnote\relax\footnotetext{
$\dagger$ Beijun Shen is the corresponding author.}$

Deep learning is a cornerstone methodology in artificial intelligence, fundamentally centered on emulating the architecture and functionality of human neural networks to autonomously learn and extract intricate patterns from data~\cite{ben2019demystifying}. In recent years, the advent of the Transformer architecture~\cite{Transformer} has catalyzed groundbreaking advancements, enabling deep learning to achieve unprecedented success across a diverse array of domains, such as computer vision~\cite{athans2018cv}, natural language processing~\cite{young2018recent}, and software engineering\cite{HouZLYWLLLGW24}.
However, the development and optimization of deep learning projects remain inherently intricate, demanding substantial domain-specific expertise. Therefore, automated deep learning code generation emerges as a promising paradigm to mitigate the challenges faced by developers, reducing barriers to the construction and deployment of deep learning models.

In recent years, pre-trained models have demonstrated exceptional performance in code generation tasks~\cite{li2022competition, liu2020multi, svyatkovskiy2020intellicode, liu2024your, dong2024self, yu2024codereval}. With the advent of large language models (LLMs), their robust generative capabilities and adaptability have established them as the dominant paradigm in this field. 
However, unlike general-purpose code, deep learning projects are characterized by their complex structures, longer functions, and a strong reliance on domain knowledge. 
Deep learning code typically encompasses more sophisticated logic, incorporating sequential steps such as data preprocessing, model construction, compilation, training, and model evaluation. These steps are highly interdependent, forming an elaborate code chain that often exceeds 300 lines. Current LLMs exhibit a significant performance degradation when generating code segments longer than 50 lines~\cite{evalLLM}, and maintaining contextual coherence across such extensive chains presents a considerable challenge.  This poses a significant obstacle for LLMs, which lack coherent contextual guidance and domain expertise during the code generation process.

Recent studies~\cite{cot2022wei} have demonstrated that the chain-of-thought (CoT) technique, which integrates intermediate reasoning steps into the prompt, can effectively guide LLMs in accurately comprehending task requirements. Moreover, the incorporation of programmatic structural information into the reasoning chain has been shown to significantly improve the performance of code generation tasks~\cite{li2023structured, autocot2022zhang, jiang2024self}.
Additionally, retrieval-augmented generation (RAG)~\cite{rag2020lewis} has emerged as a highly effective optimization strategy. This approach leverages the retrieval of relevant code snippets and supplementary domain-specific knowledge to construct enriched prompts, thereby addressing the inherent lack of domain expertise in LLMs and reducing the occurrence of hallucinations that may arise from this limitation.

Inspired by these techniques, we propose \ourname, a novel planning-guided method for deep learning code generation. Instead of presenting the reasoning process during code generation, \ourname prioritizes the formulation of a meticulously structured solution plan prior to generation. This plan acts as a global blueprint, effectively bridging the gap between user requirements with the final code output while ensuring coherence and alignment with domain-specific constraints.

We train a GPT-2 language model as a solution plan predictor using  a parallel corpus specifically constructed for this purpose.
This model generates tailored solution plans based on user requirements, offering comprehensive global guidance for code generation. To further improve the quality of the generated code, we implement two RAG strategies, namely Code RAG and Template RAG, to produce intermediate results. These results are then integrated through a comparative learning mechanism, which leverages the consistency ensured by the solution plan to effectively combine the strengths of both strategies, ultimately yielding optimized, contextually coherent, and domain-aligned deep learning code.

To evaluate the effectiveness of \ourname, we introduce \dlcodedataset, a benchmark specifically designed for deep learning code generation, curated from open-source projects. Experimental results show that \ourname outperforms existing baselines, achieving substantial improvements of 9.7\% and 3.6\% in CodeBLEU and human evaluation metrics, respectively. Remarkably, the solution predictor exhibits superior performance compared to LLMs with larger parameter scales. 
Ablation studies further emphasize the pivotal role of the proposed comparative learning mechanism and the dual RAG strategies in driving the model's success.

The contributions of our work are summarized as follows:
\begin{itemize}
    \item We propose \ourname, a novel planning-guided method for deep learning code generation. \ourname employs a solution plan predictor to generate a structured plan prior to code synthesis, offering global guidance that ensures coherence and alignment with domain-specific requirements throughout the generation process.
    
    \item We design a comparative learning mechanism that effectively integrates the strengths of Code RAG and Template RAG strategies. This mechanism enhances the quality of the generated deep learning code by simultaneously optimizing fine-grained implementation details and high-level architectural logic.  
    
    \item We construct \dlcodedataset, a benchmark dataset for deep learning code generation tasks, and compare \ourname with state-of-the-art code generation approaches using this dataset. Experimental results demonstrate that \ourname achieves significant performance improvements, surpassing baselines by a considerable margin.

\end{itemize}

\section{Related Work}
\label{sec:related}
\subsection{Generating Deep Learning Code}
Research on the generation of deep learning code remains relatively underexplored and is widely recognized as more intricate than the generation of general-purpose code. Two closely related research domains align with this objective: low-code deep learning programming and automated machine learning (AutoML).

\textbf{Low-code programming} aims to streamline the construction of artificial intelligence workflows, often through visual programming interfaces~\cite{weka}. However, proficiency in these tools still necessitates substantial time and expertise, which conflicts with the fundamental principles of low-code development~\cite{dash2023ai}. To address these limitations, Phothon Wizard~\cite{phothonwizard} integrates graphical user interfaces with dynamic code generation technologies, unifying multiple machine learning libraries and providing modular explanations. LowCoder~\cite{lowcodeai} introduces an innovative paradigm that combines visual programming with natural language programming, enabling automatic API prediction and recommendation through natural language descriptions. Text-to-ML~\cite{xu2024large} investigates the generation of deep learning code from task descriptions, however its evaluation is constrained by limited task diversity.

While these advancements have contributed to simplifying deep learning programming, they still impose certain demands on user expertise. Our research seeks to further lower the development barrier by enabling the fully automated construction of deep learning projects while accommodating a broad spectrum of complex tasks with flexibility.

\textbf{AutoML}~\cite{gu2024large} also strives to automate the deep learning workflow. AutoML-GPT~\cite{zhang2023automl} utilizes LLMs to simulate model evaluation, circumventing the need for actual execution. MLCopilot~\cite{zhang2023mlcopilot} enhances the generation of construction recommendations by extracting insights from knowledge bases, while HuggingGPT~\cite{shen2024hugginggpt} employs LLMs to orchestrate existing deep learning models for task completion. However, these studies primarily focus on identifying the most suitable model, whereas our objective is to automatically generate comprehensive deep learning code capable of executing specified tasks with minimal user intervention.

\begin{figure*}[!t]
\centerline{\includegraphics[width=1.0\textwidth, trim=0 0 0 0, clip]{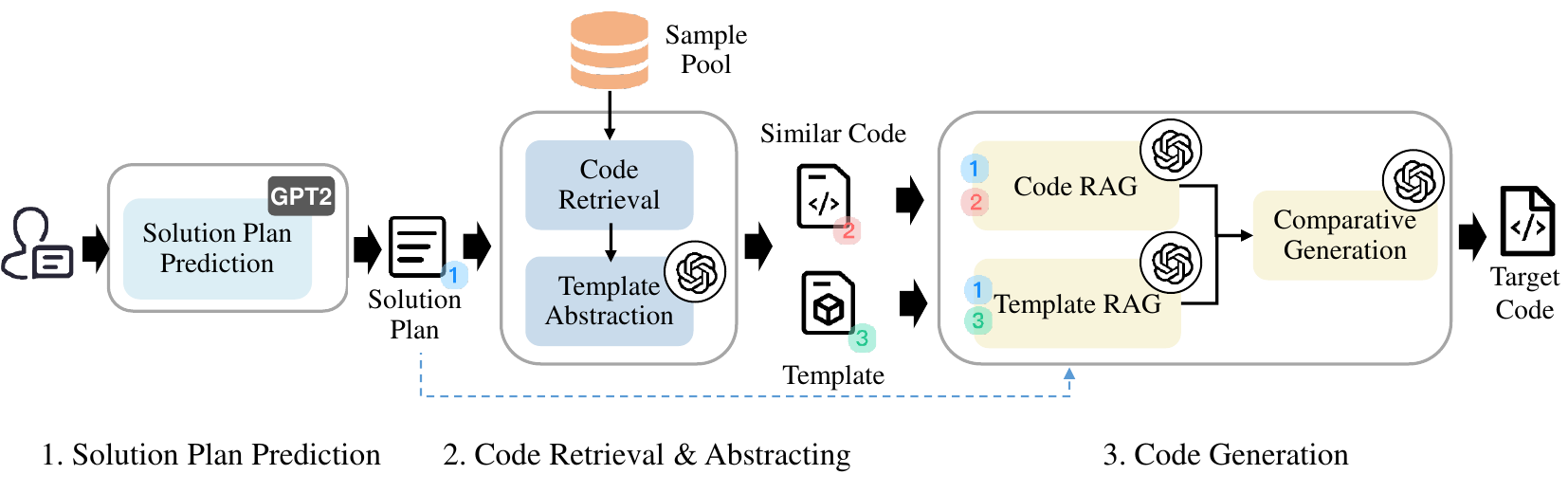}}
    \caption{Overview of \ourname} 
    \label{fig:overview}
\end{figure*}

\subsection{Chain-of-Thought Prompting}
With the rapid evolution of LLMs, Chain-of-thought (CoT)~\cite{cot2022wei}  has emerged as a pivotal technique for enhancing code generation by incorporating logical reasoning steps into prompts. 
S-CoT~\cite{li2023structured} further refines this approach by structuring intermediate reasoning steps using programmatic constructs such as sequences, branches, and loops, thereby aligning more closely with the cognitive processes of developers. Auto-CoT~\cite{autocot2022zhang} addresses the challenge of selecting high-quality examples for few-shot and manual CoT by clustering queries, identifying representative examples from each cluster, and generating corresponding reasoning chains through zero-shot prompting. Plan-and-Solve (PS)~\cite{plansolve2023wang} mitigates common issues in traditional reasoning chains, such as omitted steps and semantic misinterpretations, by first decomposing problems into smaller, manageable subproblems and then solving them in a sequential manner. 

Other research efforts harness the intrinsic knowledge of LLMs to achieve self-improvement in code generation. 
LDB~\cite{zhong2024ldb} utilizes LLM-generated evaluations of code blocks as iterative feedback to progressively optimize the output. Self-Planning~\cite{jiang2024self} employs few-shot prompting to derive solution steps from problem intents and incrementally generates code based on these plans. Similarly, SelfEvolve~\cite{jiang2023selfevolve} implements a two-stage framework, wherein the LLM first synthesizes domain-specific knowledge relevant to the problem before generating the corresponding code. 

In contrast to these studies that rely solely on the intrinsic knowledge of LLMs to construct reasoning processes, \ourname leverages a dedicated predictor trained specifically to generate deep learning solution plans. By incorporating domain-specific knowledge through fine-tuning, our approach significantly enhances the accuracy and contextual relevance of the planning process, ensuring a more robust alignment with the complexities of deep learning tasks.

\subsection{Retrieval-Augmented Generation}

Retrieval-augmented generation (RAG)~\cite{rag2020lewis} improves output quality by retrieving relevant information from external sources and integrating it with the original query. This approach is particularly effective for knowledge-intensive tasks like code generation, as it mitigates LLM challenges such as hallucinations~\cite{zhang2023siren} and outdated knowledge~\cite{jang2021towards}.

Several studies have advanced the application of RAG in code generation.
REDCODER~\cite{parvez2021retrieval} retrieves relevant code snippets from a corpus to enrich the context for code generation, while ReACC~\cite{lu2022reacc} combines token-level and semantic similarity metrics to extract pertinent code. CEDAR~\cite{cedar2023nashid} achieves state-of-the-art performance in assertion generation and program repair by leveraging embedding-based retrieval and frequency analysis. DocPrompting~\cite{zhou2022docprompting} emulates programmers' use of documentation by querying relevant fragments to assist in target code generation.

Despite these advancements, RAG-based methods remain highly dependent on retrieval quality; irrelevant, erroneous, or redundant content can significantly degrade the quality of generated code. Furthermore, existing approaches often lack effective mechanisms to bridge the gap between retrieval and generation, resulting in outputs that may not fully align with user requirements. To address these limitations, \ourname introduces dual retrieval-augmented strategies: retrieving similar code for concrete implementation references and abstracting templates to enhance generalization. A comparative learning mechanism synergizes the strengths of both strategies, while deep learning solution plans ensure semantic coherence and consistency throughout the generation process.

\section{Approach}
\label{sec:approach}
\subsection{Overview}

Given a natural language requirement \( X = x_1, x_2, \dots, x_m \), the objective of our approach is to generate the corresponding source code for a deep learning project \( C = c_1, c_2, \dots, c_n \). 
To address the challenges posed by the inherent complexity of deep learning projects, we propose \ourname, a novel planning-guided generation method. \ourname employs a fine-tuned model to predict a structured solution plan, offering global guidance for LLMs to generate comprehensive and coherent deep learning projects. 

As illustrated in Figure \ref{fig:overview}, the pipeline of \ourname consists of three key stages. First, a GPT-2 model is fine-tuned to predict a solution plan for a deep learning project based on a given natural language requirement. Second, the generated plan is leveraged to retrieve semantically analogous code samples (Code RAG) and subsequently abstract a code template (Template RAG) by utilizing an LLM. Third, a comparative learning mechanism is designed to synthesize the final code through the integration of Code RAG and Template RAG, which serve as the references for fine-grained implementation details and high-level architectural logic, respectively. 
Each step is elaborated in detail in the following subsections.

\subsection{Solution Plan Prediction}

Unlike general-purpose code, deep learning projects often adhere to consistent architectural patterns, enabling the use of structured solution plans to guide LLMs. A \emph{solution plan} encapsulates the essential components and standardized workflows intrinsic to deep learning projects, encompassing the entire pipeline from data preprocessing to model evaluation.

A solution plan typically consists of the following four core components:

\begin{enumerate}
\item \textbf{Task Category}: Identifies the primary type of machine learning task, such as Image Classification, Text Generation, or Object Detection.
\item \textbf{Dataset}: Describes the specifications of both input and output data, including their dimensionalities, formats, and other relevant attributes.
\item \textbf{Preprocess}: Outlines the sequence of transformations necessary to convert raw data into a format suitable for model input, including operations such as normalization, feature extraction, and the partitioning of datasets into training, validation, and test sets.
\item \textbf{Model Architecture}: Specifies the structural configuration of the model, including layer compositions, connectivity patterns, and the initialization and optimization of hyperparameters.
\end{enumerate}

\begin{figure}
\centerline{\includegraphics[width=0.5\textwidth, trim=0 10 0 10, clip]{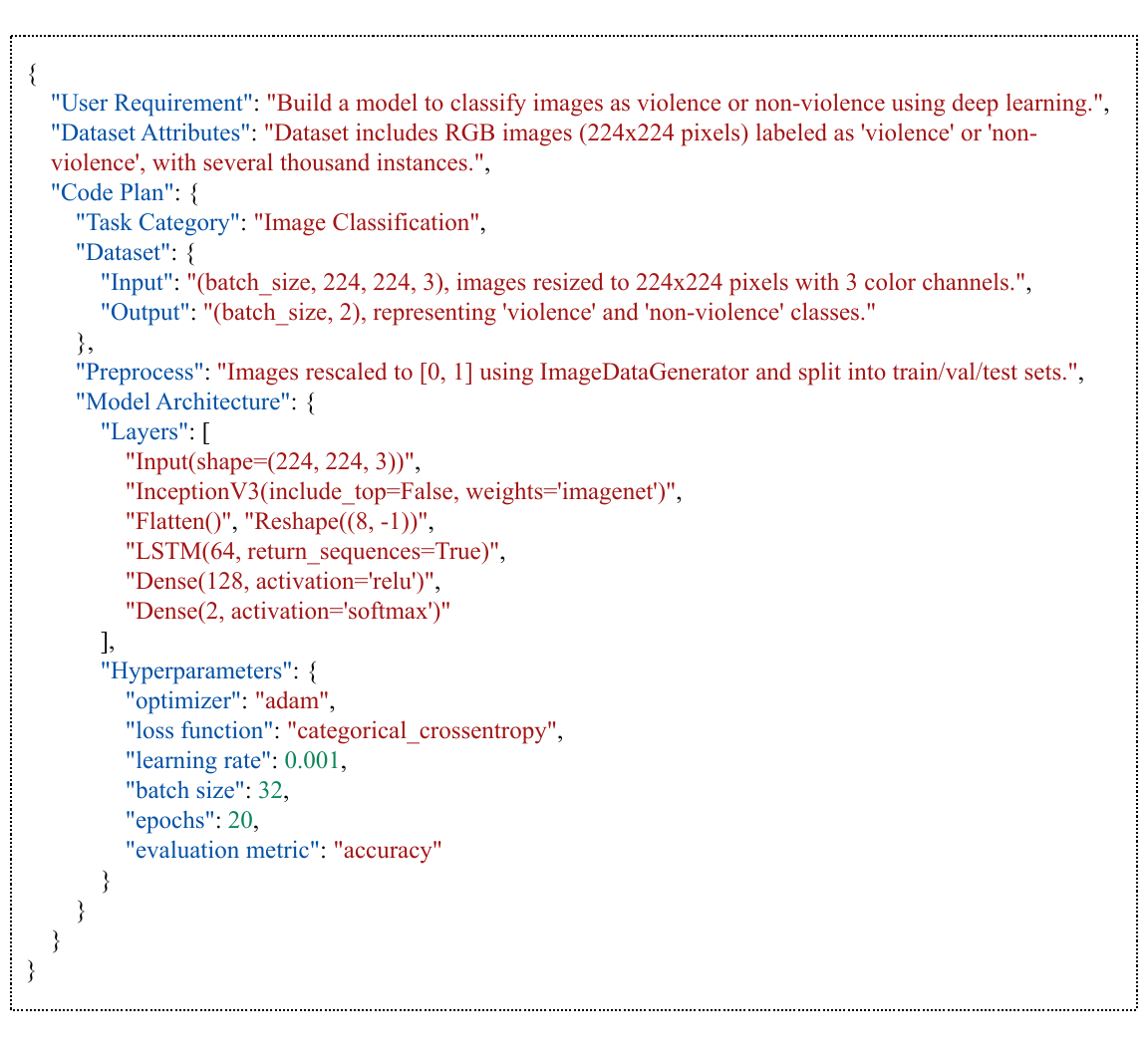}}
\caption{\fontsize{9.9pt}{0pt}\selectfont An example of a solution plan for a deep learning project} 
\label{fig:solution-example}
\end{figure}

Figure \ref{fig:solution-example} presents an illustrative example of a solution plan for an image classification task. Initially, all input images are normalized to ensure a standardized data distribution, followed by dataset partitioning facilitated by the ImageDataGenerator utility. The proposed model architecture consists of six layers,

optimized using the Adam optimization algorithm in conjunction with the cross-entropy loss function to achieve effective parameter tuning.

Given natural language requirements as input, our objective is to predict a comprehensive solution plan that functions as a structured blueprint to guide subsequent code generation. To formalize this problem, we frame it as a sequence-to-sequence learning task. We address this task by fine-tuning a pre-trained GPT-2 model, capitalizing on its robust language understanding capabilities to generate precise and contextually relevant solution plans.

\subsection{Similar Code Retrieval and Abstracting}

The generated solution plans play a pivotal role in guiding LLMs to generate deep learning projects. To maximize their utility, we employ these plans as foundational prompts and further enrich the LLM's input through the retrieval of similar code and the abstraction of high-level templates, i.e., Code RAG and Template RAG.

\textbf{Code RAG:} \ourname retrieves semantically relevant code from a pre-built code sample pool using the generated solution plan as a query. The retrieval process is conducted in two stages: first, the search scope is narrowed based on the task type specified in the solution plan; second, semantically relevant code samples are identified through similarity matching between solution plans using the BM25~\cite{robertson1995okapi} algorithm.

More specifically, the code sample pool is organized into a collection of task-specific subsets, formally represented as
\( B = R_1, R_2, \dots, R_l \), 
where each subset 
\( R_i = \{(S_{i1}, C_{i1}), (S_{i2}, C_{i2}), \dots \} \) 
consists of multiple solution-code pairs \( (S_{i}, C_{i}) \), each corresponding to a distinct category of deep learning tasks.
Given a newly generated solution plan \( S_{\text{new}} \) derived from a user requirement, we first identify the corresponding subset \( R \) according to its task category. We then compute the similarity scores between \( S \) and each plan \( S_j \) in \( R_t \) using the BM25 algorithm. The top-ranked code samples, as determined by their similarity scores, are selected to inform the subsequent code generation process.

\textbf{Template RAG:} 
The implementation details present in the retrieved code samples may introduce noise that interferes with the generative capabilities of LLMs. To mitigate this issue, \ourname synthesizes the retrieved code samples into a higher-level code template, which provides a logically coherent and domain-agnostic framework for subsequent code generation. The abstraction process effectively eliminates implementation-level variations while retaining the essential structural and functional patterns necessary for the target task.

Specifically, we utilize an LLM to extract recurring structural patterns from the top two retrieved code samples, encompassing data preprocessing workflows, model architecture definitions, and training pipeline configurations. Furthermore, we generalize the core interfaces and logical frameworks of each functional module, removing implementation-specific details such as hardcoded data formats or task-dependent configurations. These modularized frameworks are then integrated into a unified template that demonstrates high transferability and reusability across a wide range of deep learning tasks.

\subsection{Code Generation}

\begin{figure}[!t]
\centerline{\includegraphics[width=0.5\textwidth, trim=0 0 0 0, clip]{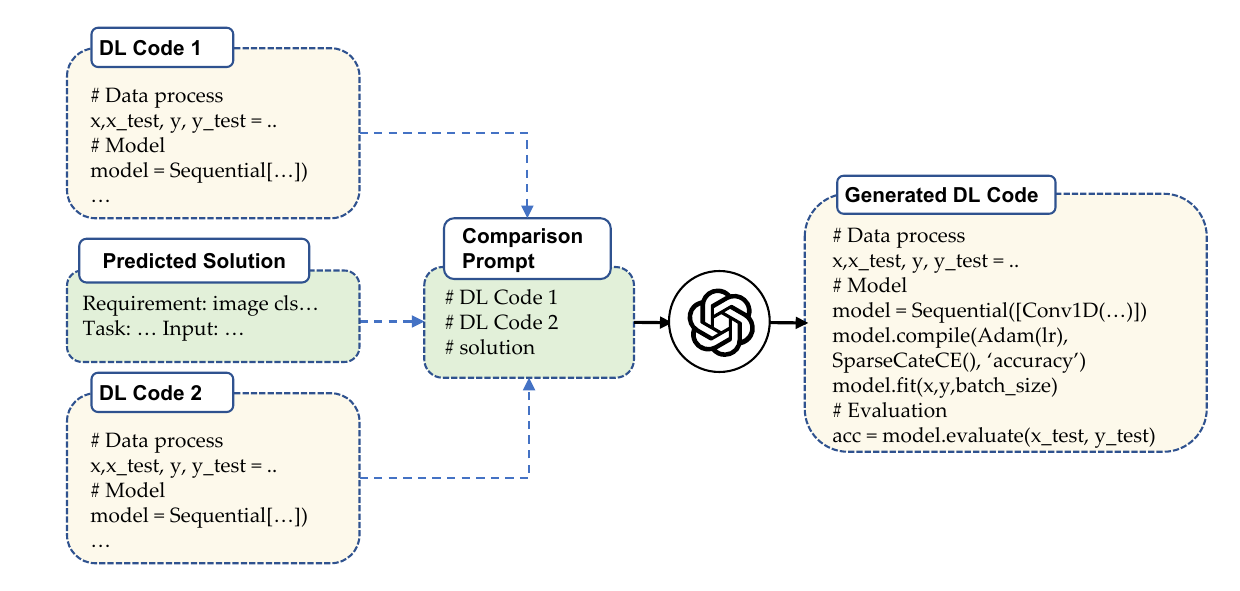}}
    \caption{An Illustration of Comparative Generation} 
    \label{fig:comparative_generation}
\end{figure}

Guided by the solution plan, \ourname utilizes a comparative learning mechanism to integrate the outputs of Code RAG and Template RAG, directing LLMs to generate the final code, as illustrated in Figure \ref{fig:comparative_generation}. 
In this framework, the solution plan provides global guidance to ensure contextual consistency, while Template RAG offers high-level structural frameworks and Code RAG delivers implementation-specific details.

Specifically, \ourname first prompts LLMs with similar code and abstracted templates to generate two code outputs, referred to as \textit{code 1} and \textit{code 2}, each featuring unique strengths and limitations. \textit{code 1}, generated via Code RAG, incorporates task-specific details and contextual information, though it may include extraneous elements unrelated to the target task, potentially compromising output quality. In contrast, \textit{code 2}, produced through Template RAG, adopts a broader perspective with a standardized framework, emphasizing structural coherence over granular details, which may lead to implementation deviations due to insufficient specificity.

To harness the complementary advantages of both RAG strategies, \ourname implements a comparative learning mechanism. 
This involves constructing comparative prompts by integrating the solution plan with dual RAG-generated code, formatted as \textit{$<$solution, code 1, code 2$>$}.
The LLM is guided to compare the two provided deep learning code outputs by utilizing the comprehensive insights from the solution plan, and subsequently identifies the optimal segments from each to inform the synthesis of the final deep learning project. 
By employing comparative learning, \ourname not only corrects potential errors but also stimulates deeper reasoning, fostering greater procedural knowledge and flexibility~\cite{LBT}. This enables the LLM to effectively integrate knowledge from diverse sources, resulting in deep learning code that is more robust, logically consistent, and comprehensive.

\section{Experimental Setup}
\label{sec:experiment}

\subsection{Research Questions}

We evaluate \ourname by addressing the following research questions.

\begin{itemize}
    \item \textbf{RQ1: How effective is \ourname in deep learning code generation?}
    We assess the performance of \ourname in generating deep learning projects and compare it against state-of-the-art baseline methods.
      
    \item \textbf{RQ2: To what extent is the deep learning code generated by \ourname compliant, practical, and idiomatic?}
    In addition to automated evaluation metrics, we conduct a human study to evaluate the intrinsic quality of the generated code, focusing on its practicality for developers and adherence to programming conventions.

    \item \textbf{RQ3: How accurate is the plan prediction model in \ourname?}
    We construct a benchmark for evaluating the prediction of deep learning solution plans and compare the quality of plans generated by \ourname against those produced directly by LLMs.
    
    \item \textbf{RQ4: What is the contribution of each component in \ourname?}
    We perform an ablation study to analyze the impact of the three core components of \ourname: Code RAG, Template RAG, and the comparative learning mechanism.
    
    \item \textbf{RQ5: How does the temperature setting in LLMs influence the performance of \ourname?}
    Given the sensitivity of LLM-generated code quality to temperature settings, we investigate the effect of varying temperature values on the overall performance of \ourname.
    
\end{itemize}

\subsection{Comparison Methods}
We compare \ourname with existing methods for deep learning code generation. 
Due to the absence of techniques specifically designed for deep learning code generation, we adapt state-of-the-art general-purpose code generation approaches, including planning-guided methods, chain-of-thought prompting, and retrieval-augmented generation, to the task of deep learning project generation. Additionally, we include direct LLM generation as a baseline for comparative analysis.
Overall, we evaluate \ourname against the following baseline methods:

\begin{itemize}
    \item \textbf{Direct}:
    A baseline approach where deep learning code is generated directly by LLMs based on user-provided requirements without additional guidance or augmentation.
      
    \item \textbf{PS} (Plan-and-Solve Prompting)~\cite{plansolve2023wang}:
    A planning-guided approach that instructs the LLM to explicitly formulate a solution plan before generating code, producing intermediate reasoning steps to guide the final output.
    
    \item \textbf{C-CoT} (Cluster-CoT)~\cite{autocot2022zhang}:
    A chain-of-thought (CoT) prompting approach that selects representative problems from each task category and employs heuristic methods combined with zero-shot CoT to generate task-specific reasoning chains. During code generation, the model utilizes the relevant CoT prompt corresponding to the task category.
    
    \item \textbf{CEDAR} (Code Example Demonstration Automated Retrieval)\cite{cedar2023nashid}:
    A retrieval-augmented approach that extracts task-relevant code examples from a codebase to construct contextually enriched prompts for code generation.
    
\end{itemize}

We replicate the results of PS, C-CoT, and CEDAR using their officially released codebases. To adapt PS for deep learning code generation, we modify its implementation to generate intermediate reasoning steps aligned with the deep learning solution framework we designed, followed by the generation of deep learning code. For CEDAR, we employ a sparse retrieval mechanism to extract the most relevant code examples from the codebase based on user requirements. To ensure fairness and consistency with our approach, the number of retrieved code examples is limited to one.

\subsection{Evaluation Metrics}
\label{metric}
We use two kinds of metrics to evaluate the quality of the generated deep learning code:

\noindent \textbf{CodeBLEU}~\cite{codebleu2020ren}:
CodeBLEU is a widely used metric specifically designed for code generation tasks. In addition to weighted n-gram matching, it also incorporates syntax tree matching and semantic data flow matching.

\noindent \textbf{Human Evaluation Metrics}:
To assess the intrinsic quality of the generated code, we define the following human evaluation metrics:

\begin{itemize}
\item \emph{Compliance}: Evaluates the extent to which the generated code fulfills the functional requirements specified in the user request.
\item \emph{Helpfulness}: Assesses the degree to which the generated code assists users in understanding and incrementally developing their complete project.
\item \emph{Idiomaticity}: Measures the adherence of the generated code to programming conventions, including readability, structure, and best practices.
\end{itemize}

All metrics are scaled from 0 to 10, with higher scores indicating superior performance. 

We assess the performance of the plan prediction model using the following metrics, encompassing both overall and field-specific evaluations:

\noindent \textbf{BLEU}~\cite{bleu}:
BLEU is a widely adopted metric for assessing the quality of machine translation, measuring the n-gram overlap between generated and reference text. In our experiments, we employ BLEU-4, which averages precision scores from 1-gram to 4-gram, to evaluate both the overall quality of predicted plans and \textit{Input} \& \textit{Output} fields.

\noindent \textbf{EM} (Exact Match)~\cite{zhang2023infere}:
EM measures the exact correspondence of the \textit{Task Category}, \textit{Loss Function}, and \textit{Optimizer} fields in the predicted plans.

\noindent \textbf{MLD} (Mean Levenshtein Distance)~\cite{petersen2023neural}:
MLD computes the average Levenshtein distance among elements within an array, serving to evaluate the \textit{Layers} field.

\noindent \textbf{MAE} (Mean Absolute Error)~\cite{zhou2021informer}:
MAE calculates the average absolute difference between predicted and actual values, used to evaluate numerical accuracy in the \textit{Hyperparameters} field.

\subsection{Datasets}

We construct the parallel corpus required for our experiments using the Meta Kaggle Code~\cite{mkc2023jim} dataset, which consists of hundreds of thousands of publicly accessible Python and R language notebooks sourced from the Kaggle platform.

Initially, we filter Jupyter Notebook files from the Meta Kaggle Code dataset, identified by the .ipynb extension. These files include complete code execution workflows, annotations, and output information. Subsequently, we apply keyword-based filtering using terms relevant to the TensorFlow framework to exclude notebooks employing traditional machine learning methods, retaining only those utilizing the TensorFlow and Keras frameworks. Duplicate entries are removed based on the content of the Code Cells. This selection process results in a final collection of 3,950 deep learning notebooks.

For these refined notebooks, we employ an LLM to extract and summarize corresponding user requirements and solution plans. As a result, we construct the following three datasets:

1) \textbf{\dlcodedataset} is designed to evaluate the quality of generated deep learning code. To prevent data leakage, we select notebook files with execution timestamps after 2024 and verify through LLMs that these files have not been exposed in their training data.

2) \textbf{\codebase} provides a curated collection of high-quality deep learning code samples. Each sample undergoes manual review to ensure its relevance and quality.

3) \textbf{\solutiondataset} comprises the remaining files, which are utilized as solution plan data for training, evaluating, and testing the solution plan predictor. 

The statistics of these datasets are presented in Table \ref{tab:dataset_statistics}. Notably, the average lines of code in these datasets is 331, significantly exceeding the 20-line average in the widely used HumanEval-X benchmark for general-purpose code generation. This highlights the complexity and scale of deep learning projects compared to general-purpose coding tasks.

\subsection{Implementation Details}
We select DeepSeek-V2.5\footnote{https://huggingface.co/deepseek-ai/DeepSeek-V2.5} as the base model for code generation. 

For the plan predictor, we employ GPT-2-large \footnote{https://huggingface.co/openai-community/gpt2-large} as the foundational model and fine-tune it on our \solutiondataset.
The model is initialized with the open-source weights provided by HuggingFace, utilizing the default BPE tokenizer and adhering to standard text length limitations. We optimize the model using the AdamW optimizer with a learning rate of 1e-3, a batch size of 16, and train it for 8 epochs.
To determine the optimal decoding strategy, we evaluate two approaches: greedy search and beam search. While greedy search offers faster inference, it often generates less coherent text in complex tasks. To achieve higher-quality and more coherent outputs while balancing computational complexity, \ourname adopts beam search as the decoding strategy. To optimize the trade-off between search space and computational efficiency, we set the \texttt{num\_beams} parameter to 3.

\begin{table}[tb]
\caption{Dataset Statistics}
\label{tab:dataset_statistics}
\begin{center}
\begin{tabular}{lcc}

\toprule
\textbf{Dataset} & \textbf{Data Size} & \textbf{Data Content} \\
\midrule
\dlcodedataset & 100/10$^{\mathrm{a}}$ & \textless user requirement, code\textgreater \\
\codebase & 40 & \textless code, solution\textgreater \\
\solutiondataset & 3300/100/400$^{\mathrm{b}}$ & \textless user requirement, solution\textgreater \\
\bottomrule
\multicolumn{3}{l}{$^{\mathrm{a}}$Automated Testing Set / Manual Evaluation Set.}   \\
\multicolumn{3}{l}{$^{\mathrm{b}}$Train / Evaluation / Test.}    \\
\end{tabular}
\end{center}
\end{table}

\section{Results}
\label{sec:results}

\begin{table*}[tbhp]
\caption{Automated Evaluation of Various Approaches in Deep Learning Code Generation}
\vspace{-3pt}
\label{tab:main_results}
\begin{center}
\begin{tabular}{l l c c c c c}
\toprule
\textbf{Model} & \textbf{Method} & \textbf{ngram} & \textbf{w-ngram} & \textbf{syntax-match} & \textbf{dataflow-match} & \textbf{CodeBLEU}        \\
\midrule
& Direct & 4.23 & 10.61 & 47.62 & 13.43 & 18.97 \\
& PS & 4.57 & 13.31 & 49.95 & 13.70 & 20.38 \\
DeepSeek-V2.5 & C-CoT & 6.33 & 14.32 & 52.05 & 18.18 & 22.72 \\
& CEDAR & 10.71 & 15.07 & 54.94 & 20.01 & 25.18 \\
& \ourname(ours) & \bf{12.81} & \bf{18.55} & \bf{56.27} & \bf{22.44} & \bf{27.52} \\
\midrule
& Direct  & 3.02 & 9.80 & 49.39 & 12.71 & 18.73 \\
& PS & 3.05 & 11.59 & 49.94 & 12.61 & 19.30 \\
GPT-4o-mini & C-CoT & 3.52 & 11.44 & 51.42 & 16.78 & 20.79 \\
& CEDAR & 8.38 & 13.85 & 54.37 & 19.30 & 23.97 \\
& \ourname(ours) & \bf{9.78} & \bf{16.15} & \bf{54.81} & \bf{22.12} & \bf{25.71} \\
\bottomrule
\end{tabular}
\end{center}
\end{table*}

\subsection{Effectiveness of \ourname (RQ1)}

We compare \ourname against baseline methods on the \dlcodedataset\  using CodeBLEU and its four sub-metrics. To understand the transferability of our approach, we also conduct experiments on GPT-4o-mini, a proprietary state-of-the-art LLM.
As shown in Table \ref{tab:main_results}, 
\ourname consistently outperforms the baseline methods across both backbone language models, achieving the highest overall CodeBLEU score.

Specifically, when evaluated on the DeepSeek-V2.5 model,  \ourname achieves a significant improvement over the best-performing baseline, CEDAR, with a 2.34-point increase in overall performance, representing a 9.3\% enhancement. Furthermore, \ourname surpasses competing methods across all four sub-metrics, demonstrating notable improvements in n-gram, w-ngram, syntax-match, and dataflow-match scores by 19.6\%, 23.1\%, 2.4\%, and 12.1\%, respectively. 

A similar trend is observed on the GPT-4o-mini model, where \ourname achieves a 7.3\% increase in CodeBLEU over CEDAR. The improvements across individual sub-metrics are as follows: n-gram (16.7\%), w-ngram (16.6\%), syntax-match (0.8\%), and dataflow-match (14.6\%). Notably, the dataflow-match and w-ngram sub-metrics exhibit the most significant absolute improvements, with gains of 2.82 and 2.3 points, respectively.

These results indicate that the deep learning code generated by \ourname not only exhibits higher textual pattern similarity but also aligns more accurately with the semantic and data flow structures of the reference code, highlighting its superior quality and robustness.

\vspace{3pt}
\noindent \textbf{Answer to RQ1}:
\ourname consistently outperforms state-of-the-art methods in deep learning project generation across all backbone LLMs, as demonstrated by higher scores in both the overall CodeBLEU metric and its four sub-metrics.

\subsection{Qualitative Evaluation (RQ2)}
To evaluate the intrinsic quality of the generated deep learning projects, we perform a human study involving experienced programmers. Four participants from author’s institution, but affiliated with different labs, are recruited through targeted invitations. All participants are postgraduate researchers specializing in artificial intelligence or software engineering, with over three years of programming experience in deep learning projects. They independently assess the generated code and provide detailed justifications for their ratings. In cases of significant score discrepancies, a collaborative discussion is facilitated to resolve differences and achieve a consensus among all annotators.

Table \ref{tab:rq1_human_eval} presents the results of the human evaluation, as scored by the annotators. Overall, \ourname achieves the highest average scores across all metrics. Specifically, in terms of both \textit{compliance} and \textit{idiomaticity}, \ourname attains the top scores of 8.30 and 7.50, outperforming the strong baseline, CEDAR, by 7.8\% and 5.6\%, respectively. This indicates that our approach excels in generating code that not only fulfills user-specified functional requirements but also adheres more closely to programming conventions. This improvement can be attributed to the guidance provided by the predicted solution plans and the augmentation of abstract code templates. 

\begin{table}[tb]
\caption{Human Evaluation Results}
\vspace{-3pt}
\label{tab:rq1_human_eval}
\begin{center}
\begin{tabular}{l@{\hspace{15pt}}cccc}
\toprule
\textbf{Method} & \textbf{Compliance} & \textbf{Usefulness} & \textbf{Idiomaticity} & \textbf{Average} \\ 
\midrule
Direct & 7.40 & 5.40 & 6.50 &  6.43 \\
PS & 7.80 & 6.00 & 6.60 &  6.80 \\
C-CoT & 7.90 & 6.20 & 6.90 &  7.00 \\
CEDAR & 7.70 & \bf{6.40} & 7.10 &  7.07 \\
\ourname  & \bf{8.30} & 6.20 & \bf{7.50} &  \bf{7.33} \\
\bottomrule
\end{tabular}
\end{center}
\vspace{-8pt}
\end{table}

In terms of \textit{usefulness}, \ourname achieves a score of 6.20, marginally lower than CEDAR's score of 6.40. This minor discrepancy may arise from CEDAR's direct incorporation of code samples retrieved from real-world projects. While its generated outputs may not always precisely align with specific functional requirements, they often provide adaptable starting points, underscoring the inherent advantage of retrieval-augmented techniques in practical code generation scenarios.

\vspace{3pt}
\noindent \textbf{Answer to RQ2}: \ourname demonstrates significant improvements of 7.8\% and 5.6\% in \textit{compliance} and \textit{idiomaticity}, respectively, while achieving comparable performance in \textit{helpfulness} when compared to the strong baseline, CEDAR.

\subsection{Performance of Plan Predictor (RQ3)}
Given the pivotal role of the plan predictor in our approach, we evaluate its performance on the \solutiondataset\  using two decoding strategies: beam search and greedy search. For comparative analysis, we select few-shot learning-based GPT-3.5-turbo and GPT-4o-mini as baseline methods.

\begin{figure}[t!]
\centerline{\includegraphics[width=0.4\textwidth, trim=0 10 0 10, clip]{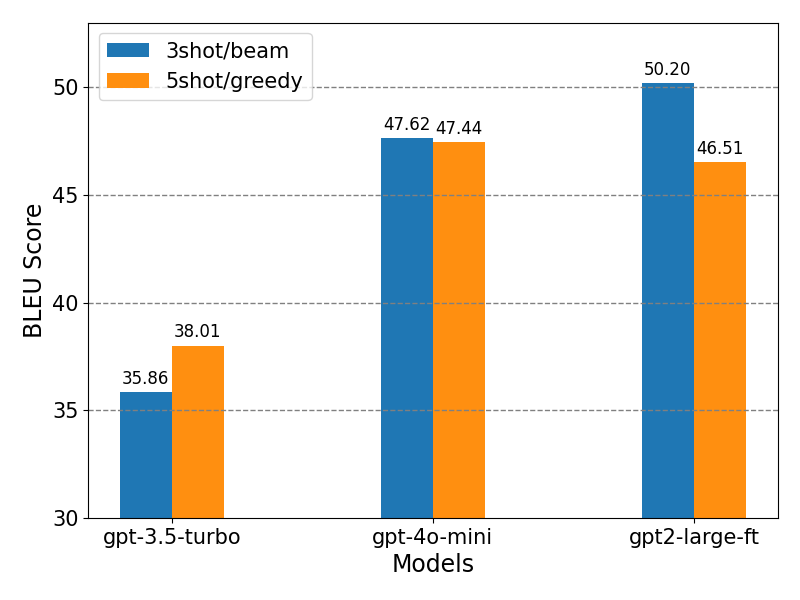}}
    \caption{Performance of Solution Plan Predictor} 
    \label{fig:rq3-overall}
\end{figure}

The results, as illustrated in Figure \ref{fig:rq3-overall}, demonstrate that the fine-tuned GPT-2-large model with beam search achieves the highest performance, highlighting the efficacy of our approach.

While GPT models exhibit strong code generation capabilities (particularly GPT-4o-mini, whose 3-shot performance surpasses that of the fine-tuned GPT-2-large model using greedy search), general-purpose LLMs still underperform in specialized deep learning code generation tasks. This limitation stems from the absence of domain-specific training and fine-tuning, emphasizing the necessity of tailored optimization for domain-specific applications to achieve optimal results.

\begin{table*}[hbtp!]
\caption{Field-level Performance of Plan Predictor} 
\label{tab:rq3_itemized}
\begin{center}
\begin{tabular}{l c c c c c c c c c c}
\toprule
\textbf{Model} & \textbf{Category$\uparrow$} & \textbf{Input$\uparrow$} & \textbf{Output$\uparrow$} & \textbf{Layers$\downarrow$} & \textbf{LR$\downarrow$} & \textbf{LF$\uparrow$} & \textbf{Optimizer$\uparrow$} & \textbf{BS$\downarrow$} & \textbf{Epoch$\downarrow$} & \textbf{Metric$\uparrow$} \\ 
\midrule
GPT-3.5-turbo(3-shot)  &
83.50\% &
23.18 &
26.71 &
35.00 &
0.20 &
40.25\% &
64.00\% &
132.99 &
71.40 &
53.75\% \\
GPT-3.5-turbo(5-shot)   &
81.25\% &
20.85 &
24.48 &
33.44 &
0.20 &
34.25\% &
63.75\% &
134.77 &
72.50 &
50.25\% \\
GPT-4o-mini(3-shot) &
88.00\% &
\bf{27.88} &
\bf{32.97} &
35.62 &
0.20 &
74.75\% &
\bf{64.25\%} &
133.04 &
70.58 &
56.75\% \\
GPT-4o-mini(5-shot)   &
89.50\% &
27.12 &
32.12 &
37.42 &
0.20 &
\bf{77.25\%} &
63.50\% &
134.00 &
71.76 &
58.50\% \\
GPT-2-large(greedy) &
86.75\% &
26.73 &
28.51 &
28.66 &
0.12 &
62.25\% &
63.25\% &
\bf{123.11} &
80.70 &
65.75\% \\
GPT-2-large(beam) &
\bf{91.75\%} &
26.90 &
32.04 &
\bf{28.39} &
\bf{0.09} &
67.00\% &
62.75\% &
128.26 &
\bf{69.14} &
\bf{66.50\%} \\
\bottomrule
\multicolumn{11}{l}{$^{*}$LR = Learning Rate; LF = Loss Function; BS = Batch size.}   \\
\end{tabular}
\end{center}
\end{table*}

To further examine the performance differences among various methods in predicting deep learning plans, we perform a field-level analysis of the generated plans. The results are summarized in Table \ref{tab:rq3_itemized}. The fine-tuned GPT-2-large model achieves superior performance across most metrics, achieving the lowest error rate of 28.39 on the critical \textit{Layers} field, significantly outperforming the two baseline LLMs. This substantial performance gap underscores the fine-tuned plan predictor's distinct advantage in accurately predicting deep learning architecture layers, a task that demands extensive domain-specific expertise.

Notably, GPT-4o-mini demonstrates exceptional performance in the \textit{Input} and \textit{Output} fields, both of which involve extensive textual descriptions. This indicates that the large-scale parameter architecture of general-purpose LLMs provides a significant advantage in handling complex text generation tasks. Conversely, the fine-tuned GPT-2-large model exhibits potential for further improvement in predicting the \textit{Loss Function} and \textit{Optimizer} fields, suggesting opportunities for enhancing domain-specific fine-tuning.

\vspace{3pt}
\noindent \textbf{Answer to RQ3}:
The fine-tuned GPT-2-large model surpasses general-purpose LLMs in solution plan prediction, delivering superior performance in both comprehensive and field-level evaluations.

\subsection{Contribution of Components (RQ4)}
To examine the contribution of the three core components, Code RAG, Template RAG, and comparative learning, to the performance of our approach, we conduct an ablation study. Specifically, we remove one or two components at a time and evaluate the performance of the resulting ablated models.

\begin{table}[t!]
\caption{Component Ablation Study}
\label{tab:rq4_ablation}
\begin{center}
\resizebox{\columnwidth}{!}{%
\begin{tabular}{l c  c  c  c  c}
\toprule
\textbf{Method} & \textbf{ngram} & \textbf{w-ngram} & \textbf{syn-match} & \textbf{df-match} & \textbf{CodeBLEU}\\ 
\midrule
\ourname & 12.81 & 18.55 & 56.27 & 22.44 & 27.52\\
- w/o Comparison \& CR  & 10.70 & 17.34 & 55.21 & 19.05 & 25.57\\
- w/o Comparison \& TR & 9.98 & 16.45 & 54.68 & 20.19 & 25.33\\
- w/o CR & 12.38 & 17.29 & 54.15 & 22.01 & 26.46\\
- w/o TR & 12.47 & 17.75 & 55.12 & 22.00 & 26.83\\
- w/o CR \& TR & 12.28 & 16.55 & 53.90 & 19.77 & 25.63\\
\bottomrule
\multicolumn{6}{l}{$^{*}$ CR = Code RAG;  TR = Template RAG.}   \\
\end{tabular}
}
\end{center}
\end{table}

The results are summarized in Table \ref{tab:rq4_ablation}. 
A significant decline in CodeBLEU scores is observed across all ablated variants. Notably, variants lacking the comparative learning mechanism exhibit the most substantial performance degradation, with score drops of 1.95 and 2.19, respectively. This underscores the critical role of the comparative learning mechanism in effectively integrating the strengths of Code RAG and Template RAG, thereby significantly enhancing the overall performance of \ourname.

The results further reveal that removing Code RAG and Template RAG leads to score reductions of 1.06 and 0.69, respectively, highlighting the individual contributions of both components to the final performance. Notably, when the outputs of these two components are replaced with those from C-CoT and CEDAR, the score drops significantly to 25.63, which is only marginally higher than the scores of variants without the comparative learning mechanism. 
This is attributed to the absence of global contextual information in the code generated by C-CoT and CEDAR, which limits the comparative learning mechanism's ability to achieve a complementary effect.

In summary, our approach employs a comparative learning mechanism to effectively integrate knowledge from Code RAG and Template RAG, collectively producing deep learning code that is more robust, logically consistent, and comprehensive.

\vspace{3pt}
\noindent \textbf{Answer to RQ4}:
Code RAG, Template RAG, and comparative learning synergistically contribute to the enhanced effectiveness of our generation technique for deep learning projects. Code RAG and Template RAG provide complementary implementation and structural insights, which are seamlessly integrated by the comparative learning mechanism to achieve robust and high-quality code generation.

\subsection{Temperature Setting (RQ5)}

We evaluate \ourname across a range of temperature settings (0.0, 0.5, 1.0, 1.5, 2.0) to investigate the impact of temperature on generation quality and determine the optimal parameter configuration.

\begin{figure}[t!]
    \centering
    \begin{subfigure}{0.24\textwidth}
        \centering
        \includegraphics[width=\linewidth]{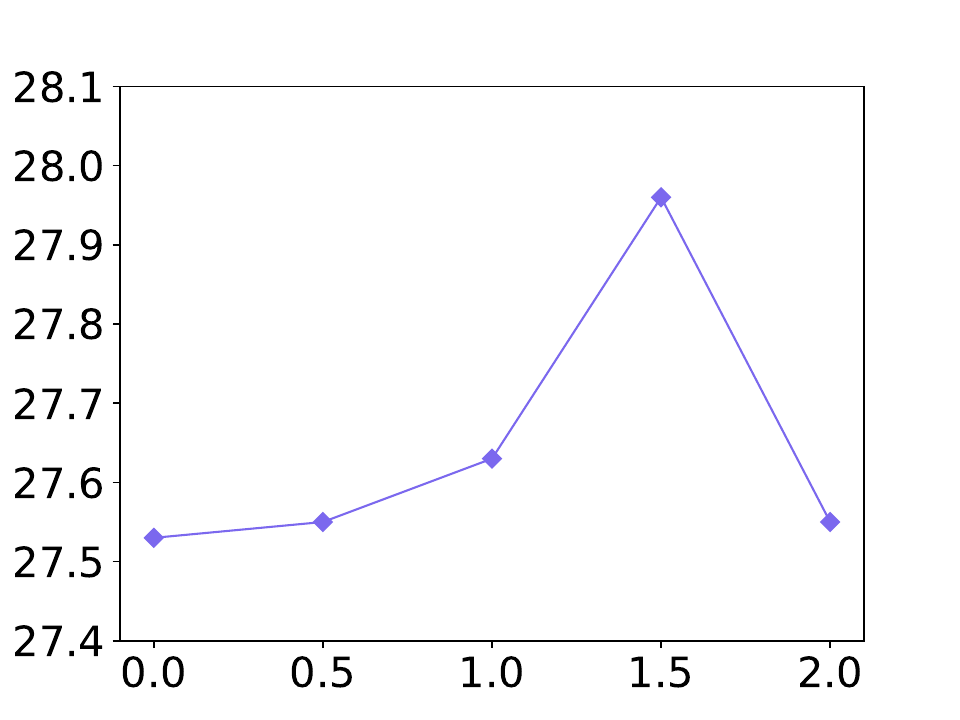}
        \caption{CodeBLEU}
    \end{subfigure}
    
    \begin{subfigure}{0.24\textwidth}
        \centering
        \includegraphics[width=\linewidth]{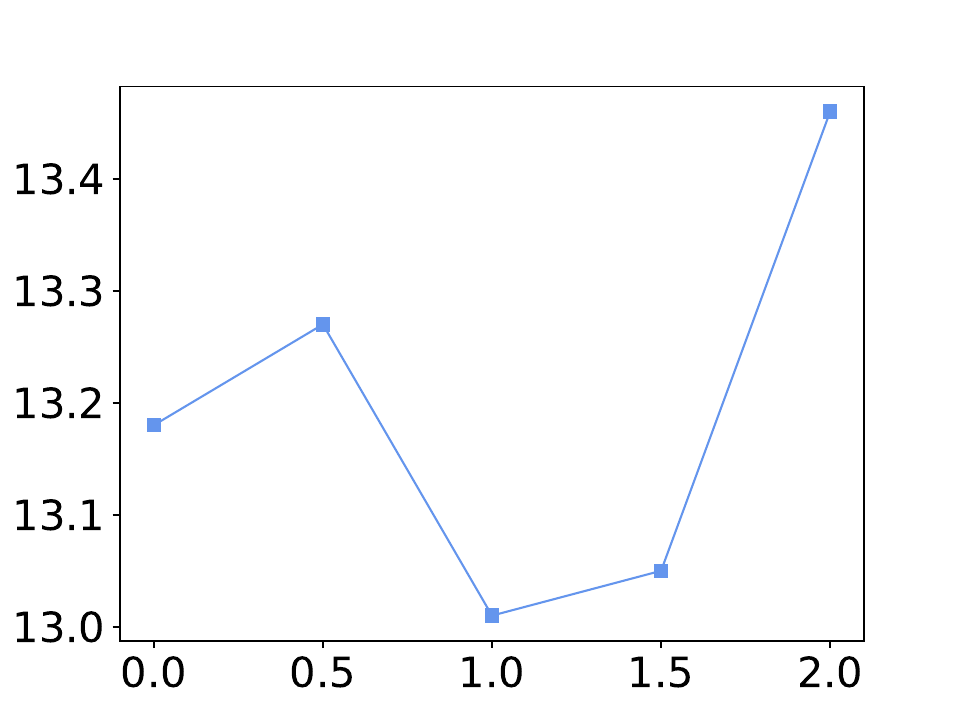}
        \caption{ngram}
    \end{subfigure}
    \hfill
    \begin{subfigure}{0.24\textwidth}
        \centering
        \includegraphics[width=\linewidth]{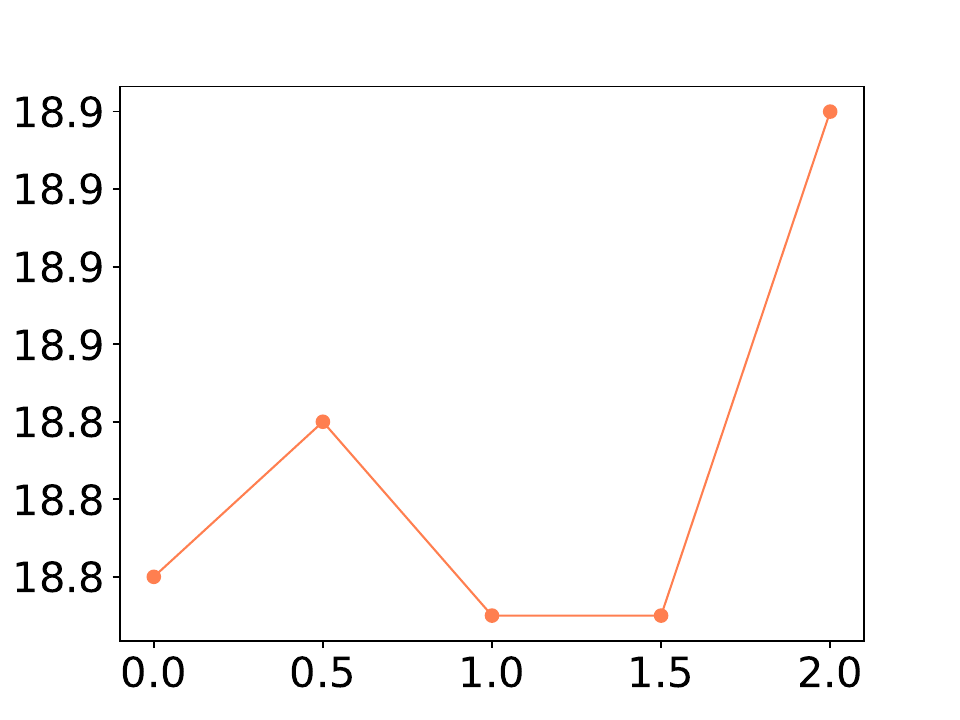}
        \caption{w-ngram}
    \end{subfigure}
    
    \begin{subfigure}{0.24\textwidth}
        \centering
        \includegraphics[width=\linewidth]{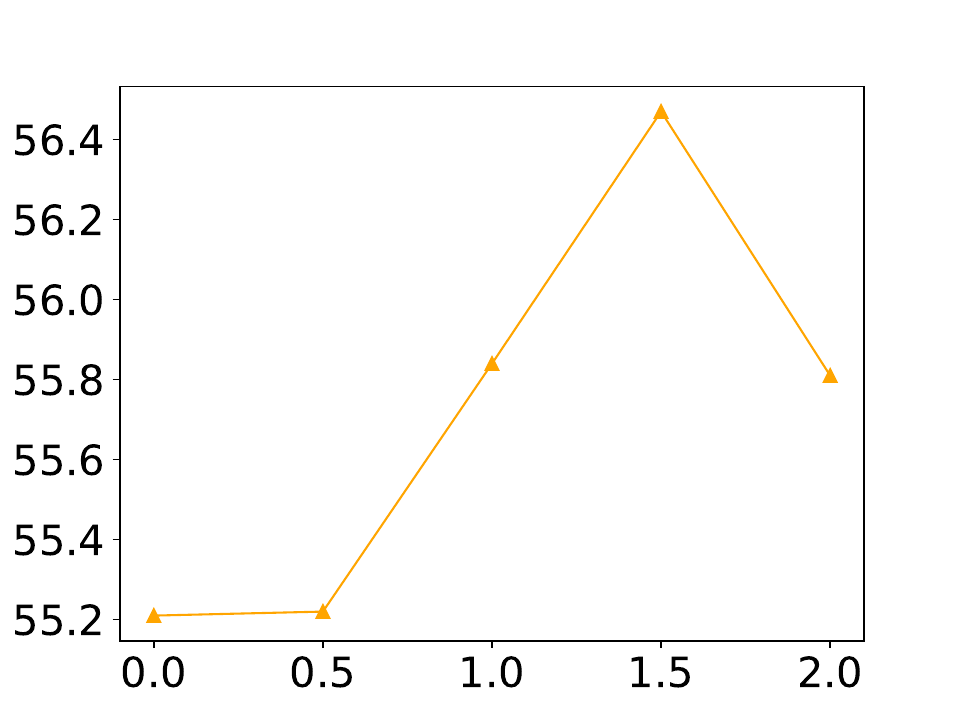}
        \caption{syntax-match}
    \end{subfigure}
    \hfill
    \begin{subfigure}{0.24\textwidth}
        \centering
        \includegraphics[width=\linewidth]{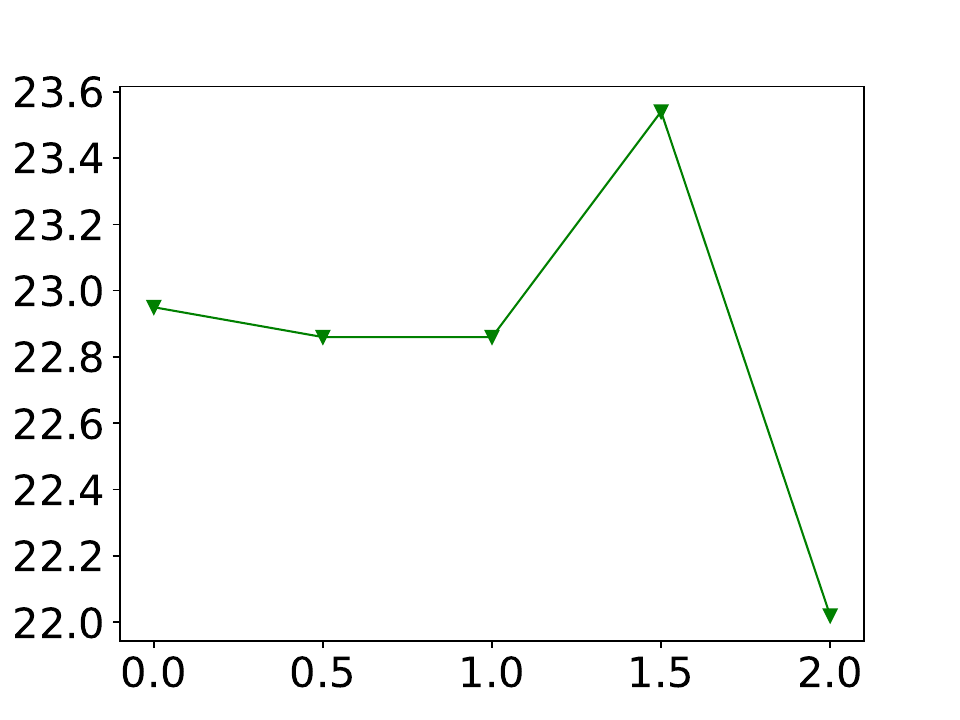}
        \caption{dataflow-match}
    \end{subfigure}
    \caption{The Trend of Scores with Temperature Changes} 
    \label{fig:rq5_temperature}    
\end{figure}

As shown in Figure \ref{fig:rq5_temperature}, the overall CodeBLEU score initially increases and then declines as the temperature rises, peaking at 27.96 with a temperature setting of 1.5. At this optimal point, the sub-metrics \textit{syntax-match} and \textit{dataflow-match} also reach their highest values, while \textit{ngram} and \textit{w-ngram} scores remain relatively lower. Considering that semantic alignment is more critical than textual similarity in code evaluation, we identify 1.5 as the optimal temperature configuration for \ourname.

It is noteworthy that this value is significantly higher than the temperature settings typically used by other baseline methods, which are often set to 0. We attribute this to the fact that a higher temperature encourages greater exploration of potential generation paths. Simultaneously, \ourname's comparative learning mechanism ensures the quality of the generated code, effectively mitigating the risk of low-quality outputs that are commonly associated with higher temperature settings.

\vspace{3pt}
\noindent \textbf{Answer to RQ5}:
The temperature setting significantly influences the quality of code generated by \ourname. Using DeepSeek-V2.5 as the base model, a temperature configuration of 1.5 yields the optimal results.

\section{Discussion}
\subsection{Why is \ourname Effective?}
We attribute the effectiveness of our \ourname to the following three factors:

\textit{1) Planning-Guided Generation:} The solution plan predictor, fine-tuned on a high-quality parallel corpus, provides domain-specific knowledge and ensures global consistency throughout the code generation process.

\textit{2) Abstract Template Augmentation:} The abstraction of general templates mitigates the sensitivity of RAG to the quality of retrieved code samples, thereby enhancing the robustness and reliability of the generated code.

\textit{3) Comparative Learning Mechanism:} This mechanism effectively integrates the strengths of dual RAG strategies, enabling error correction while stimulating deeper reasoning. This fosters greater procedural knowledge and flexibility in the generated code.

\subsection{Limitation and Threats to Validity}
We have identified the following limitations and potential threats to the validity of our method:

\textit{Data Leakage:}
\ourname relies on LLMs to generate deep learning code. As the scale of these models increases, so does the scope of their training data, raising the possibility of exposure to data from our test dataset. To mitigate this risk, we have implemented time-based filtering and knowledge detection strategies to minimize overlap between the test dataset and the training data. However, a residual risk of data leakage remains despite these precautions.

\textit{Evaluation of Generated Code:}
The effectiveness of \ourname is assessed using metrics such as CodeBLEU, Compliance, Helpfulness, and Idiomaticity. However, another widely recognized metric for evaluating code generation tasks is the test pass rate. Future work will involve developing a test-based benchmark to provide a more comprehensive evaluation of our approach's performance on real-world tasks.

\section{Conclusion}
In this paper, we present \ourname, a novel planning-guided method for deep learning code generation. \ourname fine-tunes a model on a carefully curated parallel corpus to generate solution plans that precisely align with user requirements.
These plans are are utilized to retrieve semantically analogous code samples and derive a code template. Furthermore, \ourname employs a comparative learning mechanism to integrate outputs from Code RAG and Template RAG, significantly improving the quality of the final code. We validate the effectiveness of \ourname through both automatic and human evaluations. Experimental results demonstrate that \ourname produces highly accurate and idiomatic code for deep learning projects, outperforming baseline methods by a substantial margin.

\section*{Data Availability}

Our source code and experimental data for replication are publicly available at: 
\href{https://github.com/MingshengJiao/DLCodeGen}
{https://github.com/MingshengJiao/ DLCodeGen}.

\section*{Acknowledgments}
This research is supported by National Key R\&D Program of China (Grant No. 2023YFB4503802) and National Natural Science Foundation of China (Grant No. 62102244, 62232003).

\balance
\bibliographystyle{IEEEtran.bst}
\bibliography{ref.bib}

\end{document}